\documentclass[aps,preprint]{revtex4}
\usepackage{epsfig}

\newcommand{\be}{\begin{equation}}
\newcommand{\ee}{\end{equation}}
\newcommand{\bc}{\begin{center}}
\newcommand{\ec}{\end{center}}
\newcommand{\bea}{\begin{eqnarray}}
\newcommand{\eea}{\end{eqnarray}}
\newcommand{\ba}{\begin{array}}
\newcommand{\ea}{\end{array}}

\begin{document}
\title{Some electronic and optical properties of self-assembled quantum
dots: asymmetries in a lens domain}

\author{Arezky H. Rodr\'{\i}guez and L. Meza-Montes}
\affiliation{Instituto de F\'{\i}sica, Universidad Aut\'onoma de Puebla,
Apdo. Postal J-48, Puebla, Pue. 72570, M\'exico.}

\date{\today}

\begin{abstract}
The self-assembled quantum dot with lens domain has rotational symmetry but it is
intrinsically asymmetric when the electron moves perpendicularly to its circular
base, {\it i. e.} along the rotational axis. To characterize this asymmetry, an
external electric field is applied along the positive or negative direction of the
rotational axis. We report the different Stark shifts appearing in the spectra as a
function of the field intensity for different lens domains. It is shown that for a
flat lens domain the asymmetry effects decrease, but even for very flat lenses they
can not be approximated by a cylindrical domain. Finally,  some optical properties
such as the dielectric constant and electroabsorption are studied. Signatures of the
energy spectrum reveal in these quantities. The importance of considering the proper
lens domain as long as the magnitude and direction field to tune a specific level
transition is stressed.
\end{abstract}


\maketitle

\section{Introduction}

Quantum dot (QD) structures have attracted the interest of theoretical and
experimental research due to the novel properties they exhibit as a consequence of
spatial confinement \cite{Yoffe2001,hawrylak-book}. The modern growth techniques have
led to the self-assembled quantum dot \cite{Stranski-Krastanow} with nearly a lens
shape, which is characterized by a spherical cap shape  \cite{fry00}. The electronic
properties of this so-called self-assembled quantum lens (SAQL) has been studied
previously \cite{hawrylak96-1,jpacm}. The study has included the effects of an
external electric \cite{pss,pss1} and magnetic field \cite{C-T}. Some optical
properties of such structures have also been analyzed \cite{PRB-DR}.

The lens domain is limited by the interception of two boundaries, one given by the
surface of a spherical cap with certain curvature, and the other a flat boundary.
This yields an asymmetric behaviour of the charge carriers as they are pushed toward
either surface. Consequences of this spatial asymmetry was first reported in Ref.
\cite{raymond98} for microphotoluminescence measurements and was qualitatively
reproduced in Ref. \cite{jap}.

The present work is devoted to study in more detail the asymmetry effects on the
electronic and optical properties of different lens configurations. It will be shown
that in the presence of an external electric field, the electronic and optical
properties have different behaviour depending on whether the field is along the
positive or negative direction of the axial axis.

\section{Electronic structure.}

We will consider a typical SAQL that presents a circular cross section of radius $a$
and maximum cap height $b$ with $b < a$. The lens base lies on the $xy-$plane and the
axial symmetric axis is along the $z$ axis. We want to explore the implications of
the different shapes of the lens boundaries on the physical parameters of the SAQL,
such as energy levels and wavefunctions. Thus, the SAQL will be under an external
electric field $F$ which will be along either the positive or negative direction of
the $z$ axis.

The exciton wave functions are solutions of
\be
\label{Hami}
\left[ -\frac{\hbar^2}{2m_e^*} \nabla_e^2 - \frac{\hbar^2}{2m_h^*} \nabla_h^2 - e \,
{\bf F} \, ({\bf r}_e - {\bf r}_h) - \frac{e^2}{\epsilon \left| {\bf r}_e - {\bf r}_h
\right| } \right] \Psi_{N_e,N_h} ({\bf r}_e,{\bf r}_h) = (E - E_g) \Psi_{{N_e,N_h}}
({\bf r}_e,{\bf r}_h),
\ee
where $E_g$ is the gap energy, $\epsilon$ is the dielectric constant, and $m_i^*$ is
the quasiparticle effective mass with $i = \{e,h\}$, $e$ ($h$) denoting electron
(hole).

Closed solutions of one-particle wave functions $\Psi _{N,m}$ and energy levels
$E_{N,m}$  as a function of the applied electric field and lens geometry have been
published elsewhere \cite {jpacm,jap}. Here, $N$ enumerates, for a fixed value of
$m$, the electronic levels by increasing value of energy, and $m$ is the $z$
component of the orbital angular momentum. The excitonic correction appearing in Eq.
(\ref{Hami}) has been considered in first order perturbation theory. Details can be
seen in Ref. \cite{PRB-DR}.

\subsection{Asymmetry of the ground state transition energy.}

\begin{figure}[tbp]
  \centerline{\epsfig{file=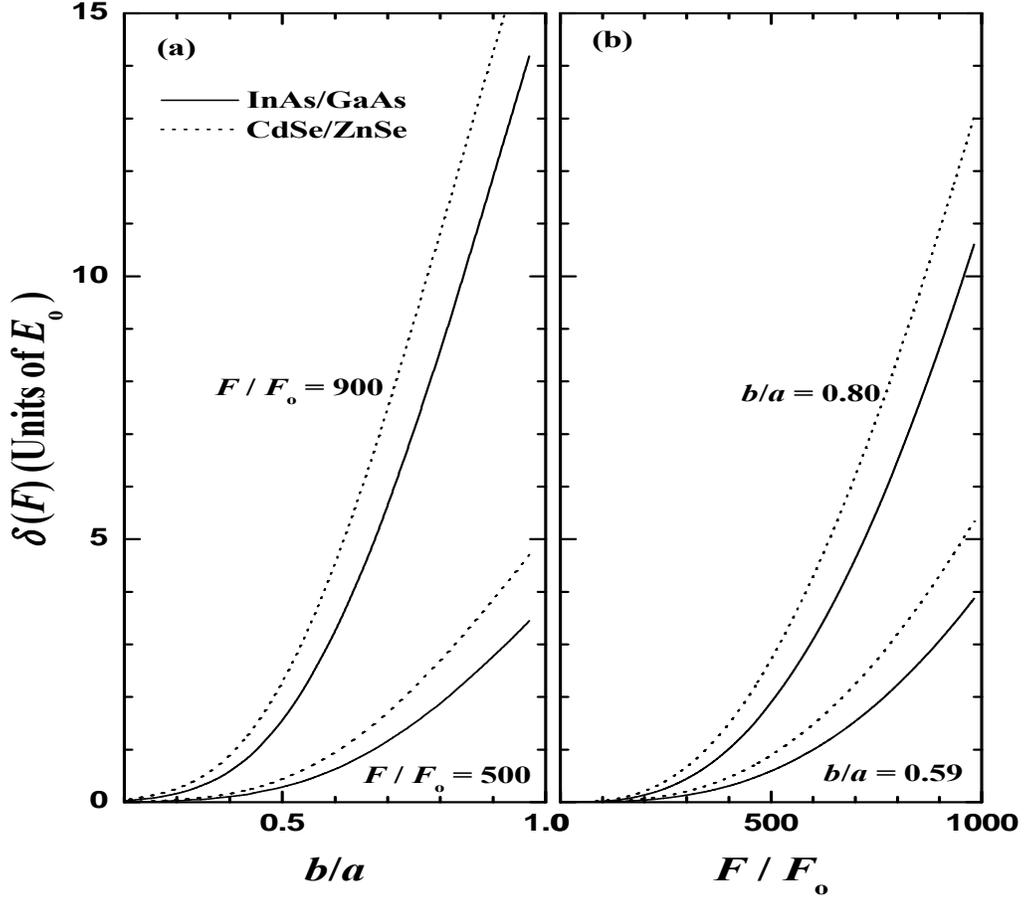,width=140mm, height=140mm, clip =}}
  \caption{Parameter $\delta$ as a function of (a) the
  lens domain deformation $b/a$ for two values of the electric field, and
  (b) the absolute value of the electric field $F$ for two lens configurations.
  For solid line the InAs parameters were used whereas the dotted line refers to CdSe SAQDS.}
  \label{figDeltaEt}
\end{figure}

In order to explore the effect of spatial asymmetry on the lens spectra, we have
defined for a given material the parameter
\be
\delta(F) = E_T(+F) - E_T(-F),
\ee
where $E_T$ is the ground state transition energy. For an electric field of magnitud
$F$ the parameter $\delta(F)$ measures the difference between the transition energies
for the fields along the positive direction $+F$ and the negative direction $-F$.

In Fig. \ref{figDeltaEt} we show (a) the parameter $\delta(F)$ as a function of $b/a$
for two values of the electric field and dots different materials,  InAs/GaAs and
CdSe/ZnSe. The unit of energy is $E_o = \hbar^2 / 2 m_e \, V_o^{2/3}$, where $m_e$ is
the free electron mass and the unit of electric field is $F_o = E_o / e \,
V_o^{1/3}$. We have used $m^*_e / m_e = 0.023$ and $m^*_{hh} / m_e = 0.34$ for InAs
\cite{emp99} whereas $m^*_e / m_e = 0.11$ and $m^*_{hh} / m_e = 0.44$ for CdSe
\cite{martin}.

The parameter $V_o$ is the volume of the lens domain. It is related to the
deformation $b/a$ and the radius $a$ of the domain through the expression $V_o = 1/2
\pi a^3 (b/a) ( 1 + 1/3 (b/a)^2 )$. Thus, the lens domain can be defined by three
parameters: the volume $V_o$, the radius $a$ and the deformation $b/a$, but only two
of them are independent. The confinement regime is determined by the volume $V_o$ of
the domain and the deformation $b/a$. If one of the them decreases, the energy levels
increase. We want to study the effects on the energy levels by changes of the lens
domain but not in terms of the confinement caused by the volume variation. Therefore,
in Fig. \ref{figDeltaEt}(a) the lens volume is kept fixed as $b/a$ is modified.

In Fig. \ref{figDeltaEt}(b) the values of $\delta$ are shown, now as a function of
the absolute value of the electric field for two different lens domains given by the
indicated values of $b/a$. Similarly to Fig. \ref{figDeltaEt}(a), the volume $V_o$
has been fixed.

It can be seen in both Figures that as  $b/a$ decreases, so does the value of
$\delta$. This could be expected due to the fact that, as the lens domain becomes
flatter, the two boundaries turn out to be more similar and then the asymmetry along
the axial axis ``is reduced''. Thus, higher field intensities $F$ are necessary for
increasing the value of $\delta$ for small $b/a$ ratio. Notice also that for the CdSe
SAQL, the values of $\delta$ in both Figures are higher compared to those of the InAs
dot. This effect is produced by the effective mass which in the CdSe material is
higher for both the electron and heavy hole. Then the energy of the ground state
transition is lower than the InAs case, and the contribution of the electric field to
energy is higher that in the InAs case for a given value of $F$.

Higher states are less affected by the field because its energy contribution is then
relatively lower \cite{pss}. Then, lower values of $\delta$ would be expected for
both materials in these cases. Therefore, for a flat enough lens domain ($b/a\ll 1$)
and not too high values of the electric field, the asymmetry in the values of the
ground and the first excited states transition energies could be neglected.  This
suggests to employ simpler geometries, as we now discuss.

\subsection{Comparison between lens and cylindrical domains.}

\begin{figure}[tbp]
  \centerline{\epsfig{file=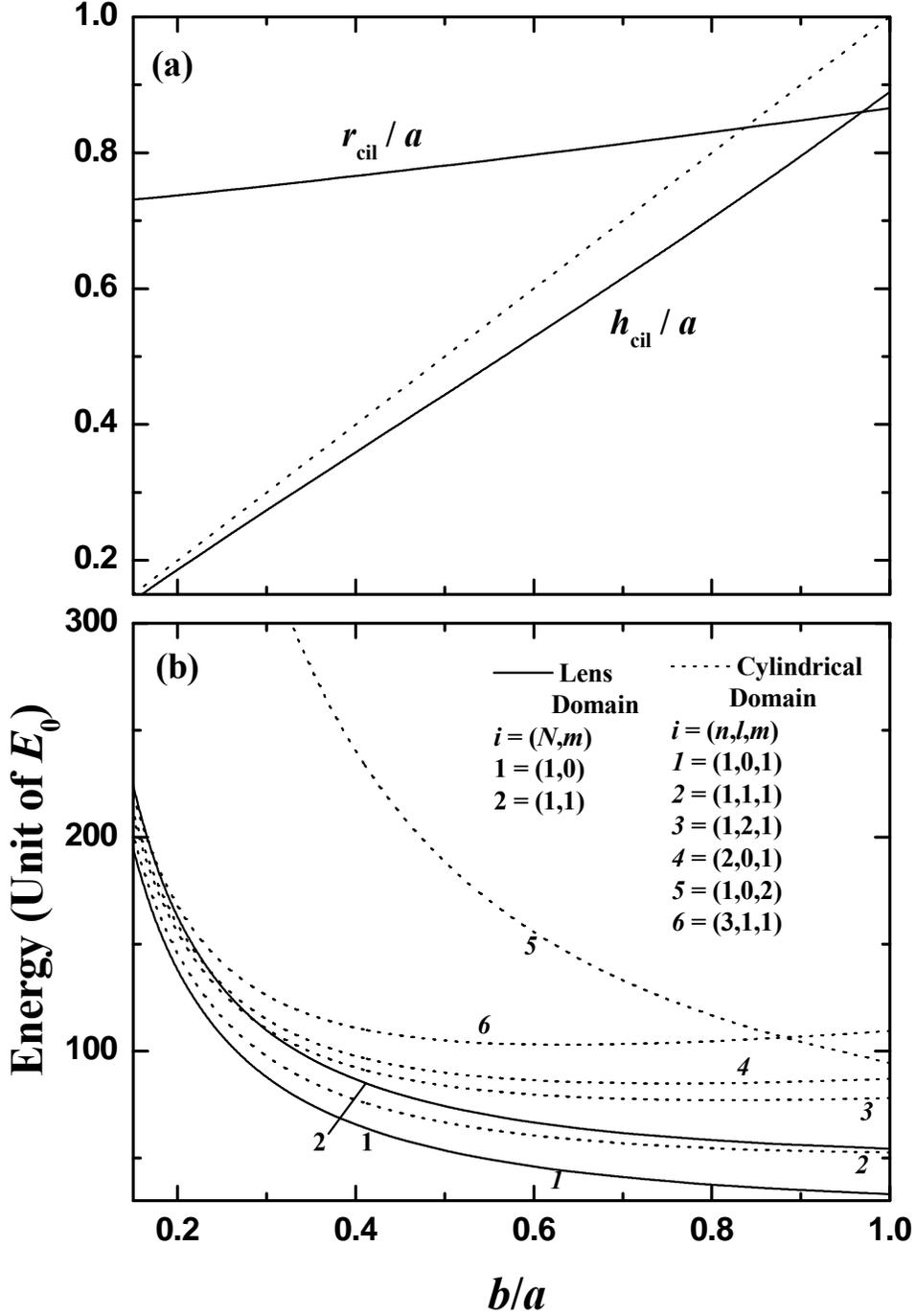,width=140mm, clip =}} 
  \caption{(a) The behaviour of the $r_{cil}$ and the $h_{cil}$ as a function
  of $b/a$ for a cylindrical configuration which fulfills conditions (1) and (2).
  (b) First two energy levels of the lens domain and first six energy levels
  of the cylindrical domain as a function of $b/a$. Quantum numbers of the
  levels are also shown in the Figure. See text.}
  \label{figCilLente}
\end{figure}

Due to the results obtained in the previous section, one could ask if a strongly
deformed lens ($b/a \ll 1$) could be approximated by a cylindrical domain with the
same volume $V_o$, radius $r_{cil}$ and height $h_{cil}$. The cylinder parameters
should be selected in such a way that the cylinder shape resembles the lens domain.

We have now the problem of selecting properly the values of $r_{cil}$ and $h_{cil}$
to obtain energy levels the more similar to those of the lens. It is also necessary
to take into account that the energy levels in the cylindrical domain are given by
\be
\label{Ecil}
E^{(cil)}_{n,l,m} = \frac{\hbar^2}{2 m^*} \left( \frac{\mu^2_{n,l}}{r_{cil}^2} +
\frac{m^2 \pi^2}{h_{cil}^2} \right),
\ee
where $\mu_{n,l}$ is the $n$-th zero of the Bessel function $J_l$ with $l=0,1,2,...$
and $m=1,2,...$  the angular momentum quantum numbers due to the rotational symmetry
of the cylinder.

We impose two conditions for any values of $b/a$: (1) the cylinder and lens domains
have the same volume $V_o$ which is kept constant, and (2) the ground electronic
state of both domains geometries are the same. This allow us to find suitable values
of $r_{cil}$ and $h_{cil}$ and to find out if the first excited states of the
cylinder are rather close to those of the corresponding lens.

There are two possible cylinder configurations which fulfill the former conditions
for a given value of $b/a$. One of them consists of a very high cylinder with
$h_{cil}\gg b$ and small $r_{cil}$. This configuration is not considered here,
instead we consider the other one since it resembles the lens domain. In Fig.
\ref{figCilLente} (a) we can see the behaviour of $r_{cil}$ and $h_{cil}$ as a
function of $b/a$, for a cylinder domain which fulfills the conditions (1) and (2).
The ratio $r_{cil}/a$ is almost constant and we have $r_{cil}/a < 1$. Contrary, the
ratio $h_{cil}/a$ strongly decreases with $b/a$. The dotted line maps directly the
values of  $b/a$ onto the vertical axis. Comparing this dotted line with $h_{cil}/a$
we conclude that $h_{cil}/a < b/a$ for the shown range of $b/a$. Consequently, we can
see that $h_{cil}$ is also smaller than the lens height $b$ and for $b/a \rightarrow
0$ we have that $h_{cil} \rightarrow b$.

In Fig. \ref{figCilLente} (b) we have plotted using continuous lines the first two
energy levels of the lens domain as a function of $b/a$. We have included also in
dotted lines the first six energy levels of the cylindrical configuration using Eq.
(\ref{Ecil}) with parameters $r_{cil}$ and $h_{cil}$ shown in Fig.
\ref{figCilLente}(a). In this case, the energy unit is $E_o = \hbar^2 / 2 m^*_e \,
V_o^{2/3}$, with $m^*_e$ the electron effective mass. Then, the plot shows the
universal behaviour of the energy levels of the lens and cylindrical domain for any
material having effective mass $m^*$. The quantum numbers of the levels are shown in
the Figure.

To analyze Fig. \ref{figCilLente} it is important to take into account that as the
lens parameter $b/a$ decreases the radius $a$ increases, for a given value of the
volume $V_o$. The behaviour of $r_{cil}$ and $h_{cil}$ in Fig. \ref{figCilLente} (a)
and Eq. (\ref{Ecil}) suggest that the cylinder levels of Fig. \ref{figCilLente} (b)
can be approximated for $b/a \ll 1$, by the expression
\be
E^{(cil)}_{n,l,m} \approx \frac{\hbar^2}{2 m^*} \frac{m^2 \pi^2}{h_{cil}^2},
\ee
showing why cylinder levels with  same $m$ tend to collapse in the $b/a \ll 1$
regime. This contrasts with the behaviour of the energy levels of the lens domain
which approach each other but remain separated. See Fig. \ref{figCilLente} (b).

It can be concluded then that the first excited levels of a flat lens can not be
approximated by those of a cylindrical domain with same volume and ground state
energy.

\section{Lens domain: influence of asymmetry on the optical pro\-per\-ties.}

The asymmetry found in the energy levels for positive and negative electric field is
not the only possible effect of the domain asymmetry along the axial axis. The
wavefunctions are also affected by the different curvatures of the boundaries,
depending on the direction of the electric field. It also give rise to the asymmetry
in other physical properties. See for example Fig. 2 in Ref. \cite{jap}, where
differences in the carrier polarizations and the oscillator strengths were also
reported. One way to study the combined influence of asymmetry on the shifts of the
transition energy and the oscillator strengths is through the dielectric constant.

\begin{table}
\label{table1}
\begin{center}
\begin{tabular}{cc}
\hline\hline
Parameters & InAs \\ \hline
$E_g$ (eV) & 0.45$^a$ \\
$m_{e}^{\ast }/m_0$ & 0.023$^a$ \\
$m_{hh}^{\ast }/m_0$ & 0.34$^a$ \\
$m_{lh}^{\ast }/m_0$ & 0.027$^a$ \\
$\Delta E_c$ (\%) & {40\%}$^a$ \\
$\Delta E_v$ (\%) & {60\%}$^a$ \\
$P^2/m_{0}$ (eV) & 10.0 $^c$ \\
$\gamma_{hh}$ (meV) & 3 \\
$\gamma_{lh}$ (meV) & 5 \\ \hline\hline
\end{tabular} \\
\medskip
$^a$ Ref. \cite{emp99} \\
$^b$ Ref. \cite{martin} \\
$^c$ Ref. \cite{landolt}
\caption{Parameters used in the calculation of the dielectric constant.}
\end{center}
\end{table}

In the following, we analyze the imaginary part of the dielectric constant for the
case of InAs/GaAs SAQLs by comparing two lens configurations. Due to the axial
symmetry of the SAQL, the interband selection rules correspond to excitonic branches
such that $m_c - m_v = 0$, where $m_c (m_v)$ is the $z$-projection of the angular
momentum in the conduction and valence band respectively. We have only considered
incident frequencies in the range below the band offset, according to the material
parameters listed in Table \ref{table1}, where $m^*$ is the effective mass of the
particle in the conduction $c$ or valence bands $hh$ and $lh$, $P$ is the interband
optical matrix element between conduction and valence bands and $\gamma_{N_c,N_{hh}}$
is the broadening parameter of the Lorentzian function. The value $E_g = 1.51$ eV for
GaAs has also been used \cite{emp99}. The theoretical formalism to calculate the
dielectric constant in SAQLs was given in Ref. \cite{PRB-DR}.

\begin{figure}[tbp]
  \centerline{\epsfig{file=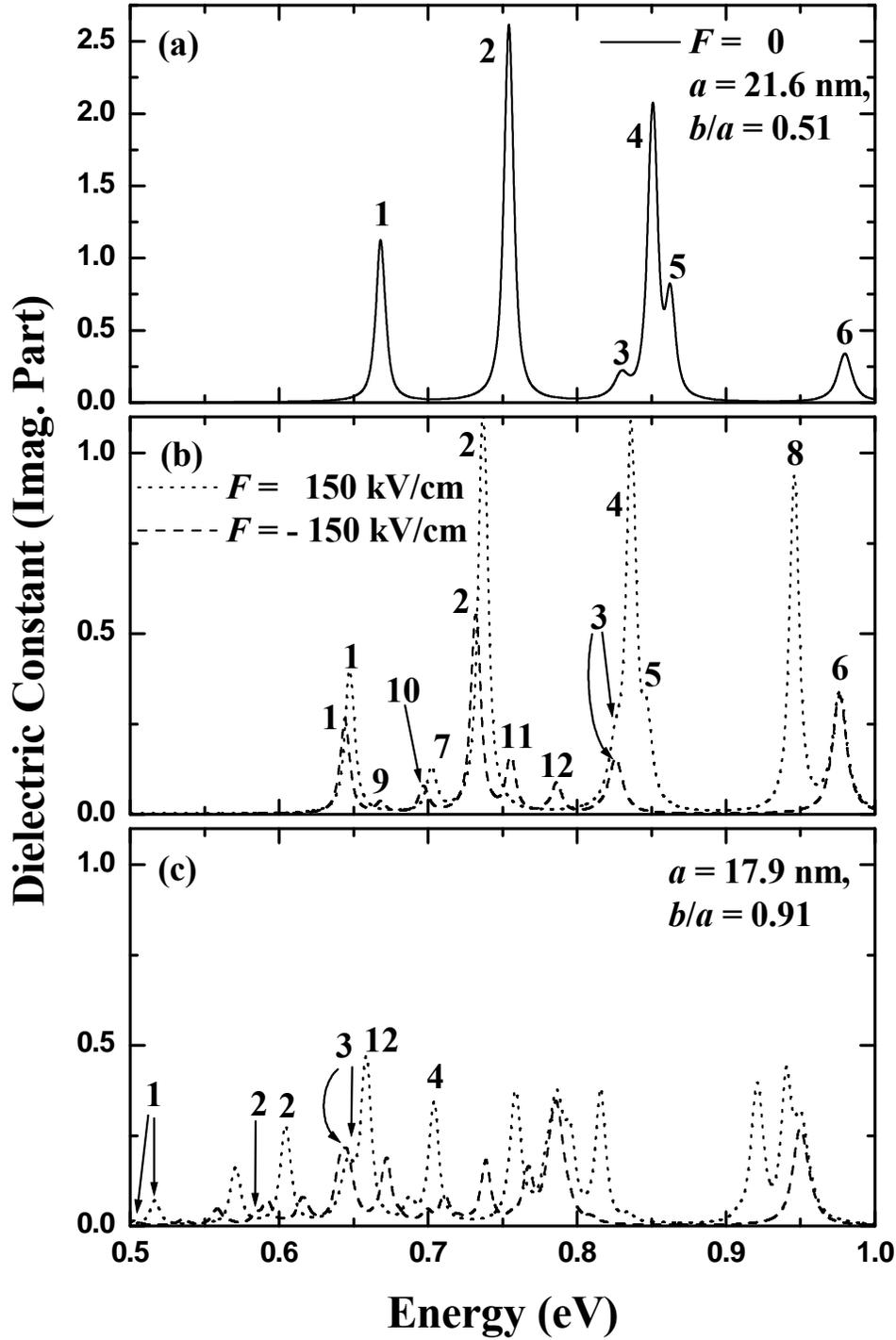,width=140mm, clip =}} 
  \caption{Dielectric constant for InAs/ GaAs SAQLs. (a) $a=21.6$ nm and $b=11.0$ nm.
  (b) Same configuration as in (a). (c) $a=17.9$ nm and $b=16.3$ nm. Both
  configurations have the same volume. Solid line represents the case with $F=0$,
  while dotted (dashed) line the case $F=150 (F=-150)$ kV/cm.}
  \label{figInAsCD2}
\end{figure}

Figure \ref{figInAsCD2} shows the dielectric constant for a SAQL of InAs embedded in
a GaAs matrix for an incident light propagating along the $z$-axis and polarized in
the plane of the SAQL base. The SAQL in Fig. \ref{figInAsCD2} (a) and (b) has a
radius $a=21.6$ nm and height $b=11.0$ nm, while in Fig. \ref{figInAsCD2} (c) the
configuration has the same volume but  $a=17.9$ nm and $b=16.3$ nm. Three values of
the electric field are considered: $F=0$ (solid line) in Fig. \ref{figInAsCD2} (a)
and $F=150$ kV/cm (dotted line) and $F=-150$ kV/cm (dashed line) in (b) and (c). It
is not shown the complete range of the band offset, but only an energy range relevant
for discussion where the first transition takes place. In Table \ref{table2} the
quantum numbers for the transitions in Fig. \ref{figInAsCD2} are denoted by
($N_c,N_v;m$) where $N$ enumerates the electronic energies by increasing value, for a
fixed value of $m$. Moreover, the numbers ($N_c,N_v;m$) indicate that the transition
takes place from level $N_v$ of the valence band ($hh$ or $lh$) to level $N_c$ at the
conduction band. Also specified are the oscillator strengths at cero, -150 and 150
kV/cm. The symbol ``-'' means that the transition does not occur because the
corresponding level at the conduction band is not present.

\begin{table}
  \centering
  \begin{tabular}{cccccc} \hline\hline
  Label & Quantum & Valence & \multicolumn{3}{c}{Oscillator Strength}  \\
        & Numbers & Band & at & $F$(kV/cm) & =   \\
        & ($N_c,N_v;m$) & & 0 & -150 & 150 \\ \hline
  1 & (1,1;0) & $hh$ & 1.0 & 0.465 & 0.579 \\
  2 & (1,1;1) & $hh$ & 1.0 & 0.450 & 0.640 \\
  3 & (1,1;0) & $lh$ & 1.0 & 0.977 & 0.977 \\
  4 & (1,1;2) & $hh$ & 1.0 & - & 0.697 \\
  5 & (2,2;0) & $hh$ & 1.0 & - & 0.548 \\
  6 & (1,1;1) & $lh$ & 1.0 & 0.981 & 0.982 \\
  7 & (1,5;0) & $hh$ & 0.0 & 0.004 & 0.344 \\
  8 & (1,1;3) & $hh$ & 0.0 & - & 0.745 \\
  9 & (1,2;0) & $hh$ & 0.0 & 0.160 & $<$ 0.05 \\
  10 & (1,4;0) & $hh$ & 0.0 & 0.262 & $<$ 10$^{-3}$ \\
  11 & (1,2;1) & $hh$ & 0.0 & 0.24 & $<$ 0.06 \\
  12 & (1,4;1) & $hh$ & 0.0 & 0.182 & $<$ 10$^{-2}$ \\\hline\hline
  \end{tabular}
  \caption{Data corresponding to the transitions labelled in Fig. \ref{figInAsCD2} (a) and (b).}
  \label{table2}
\end{table}

Comparing Figs. \ref{figInAsCD2} (a) and (b) it is clearly seen the energy red-shift
of the transitions 1 to 6  (Stark effect) at $F=\pm 150$ kV/cm respect to their
positions at $F=0$. Nevertheless, in Fig. \ref{figInAsCD2} (b) it can be seen that
these displacements are not quantitatively the same for positive or negative field.
For example, the transition peaks labelled by 1 and 2 do not appear at the same
energy leading to $\delta (F)\neq 0$ for both cases.

On the other hand, the oscillator strengths for transitions at $F=0$ are in general
higher than their corresponding values for $F=\pm 150$ kV/cm due to the mixture of
states caused as the electric field is turned on. Notice that in Fig.
\ref{figInAsCD2} (a) the scale of the dielectric constant is twice the ones used in
Figs. \ref{figInAsCD2} (b) and (c). Furthermore, transition 6 is originated in a
ligth-hole valence band and comparing it in Figs. \ref{figInAsCD2} (a) and (b) we can
conclude that the transitions for the $lh$ band are less affected by the electric
field, due to the lower effective mass than in the case of the $hh$-valence band.

It can also be seen that the transitions from valence to conduction bands with $m
\neq 0$ are relatively stronger than those with $m=0$, for a given value of the
electric field. Compare for example, transition 2 which has $m = 1$ (see Table
\ref{table2}) with transition 1 with $m=0$. It is a consequence of the excitonic
breaking of degeneracy. In the case of a transition with $(N_c,N_v,m=0)$ the
excitonic couple does not break the degeneracy. However, for a transition with
$(N_c,N_v,m\neq 0)$, a three-fold level splitting is obtained. See details in Ref.
\cite{PRB-DR}. The splitting levels remain very close each other causing closely
spaced peaks which give rise to an enhanced dielectric signal.

Forbidden transitions at cero field are allowed at $F\neq 0$ and some of them are
tuned at $F=\pm 150$ kV/cm in different ways, according to the direction of the
applied field. See for example transitions 7, 9, 10, 11 and 12. In the case of
transition 9, its oscillator strength at $F=150$ kV/cm is so small that does not
yield a visible peak, but this transition is noticeable for $F=-150$ kV/cm. On the
other hand, transition 8 only appears at $F=150$ kV/cm because the corresponding
level does not exist at the conduction band for $F=-150$ kV/cm.

In Fig. \ref{figInAsCD2} (c) the domain has the same volume as in Fig.
\ref{figInAsCD2} (a) and (b), but now the parameter $b/a = 0.91$. The transitions
take place this time at lower energies, as can be seen in the Figure. This shift is
due to the softer confinement, for the lens has a shape almost semi-spherical. For
the same reason, the effects of the electric field are reinforced, see Fig.
\ref{figDeltaEt}. Thus, the mixing of the wavefunctions becomes stronger and, as a
consequence, the oscillator strengths are more homogeneously distributed along all
the possible transitions giving rise to rich peaks distributions with quantum numbers
$N_v \neq N_c$. For the same reason, the peak heights are smaller compared to those
of Fig. \ref{figInAsCD2} (b). Differences in the energies where transitions 1 and 2
occur are bigger than in Fig. \ref{figInAsCD2} (b), in agreement with the result
obtained in Fig. \ref{figDeltaEt}.

\begin{figure}[tbp]
  \centerline{\epsfig{file=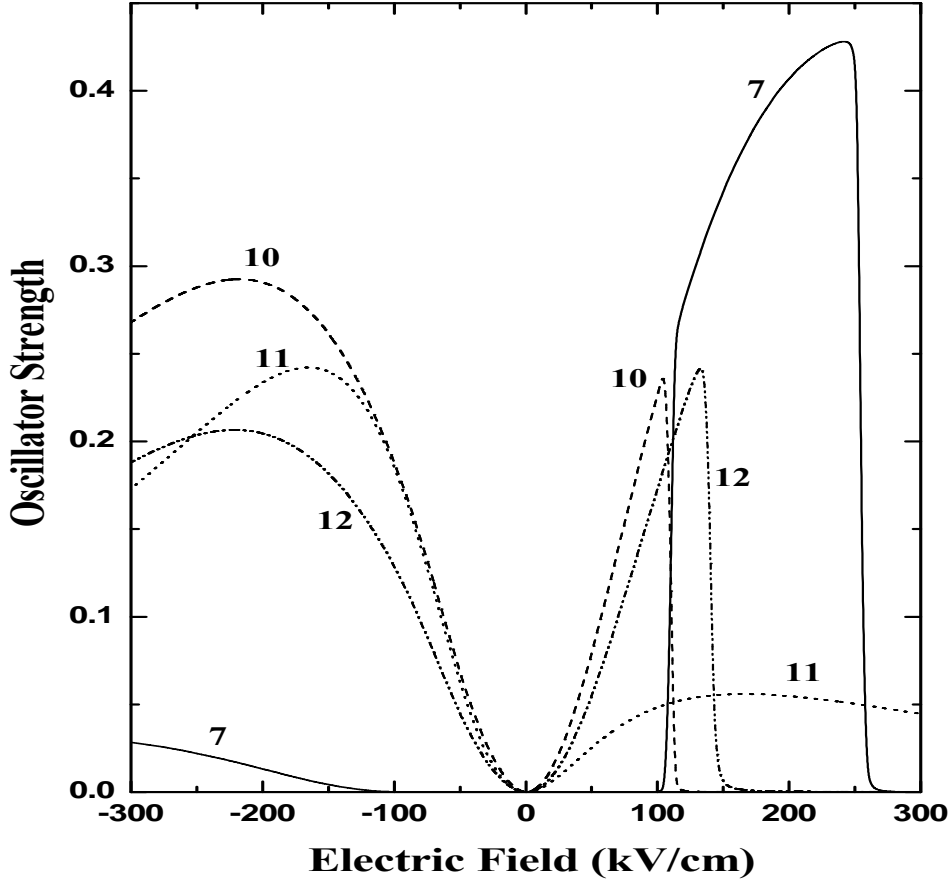,width=140mm, height=140mm, clip =}}
  \caption{Oscillator strength as function of the electric field for
  electron-hole transitions 7, 10, 11 and 12 which correspond to Fig. \ref{figInAsCD2}.}
  \label{figInAsOS}
\end{figure}

For a better understanding of the peaks distribution in the dielectric constant, in
Fig. \ref{figInAsOS} the oscillator strengths for transitions 7, 10, 11 and 12 are
plotted as a function of the electric field. We can see a strong variation of the
oscillator strength for transitions 7, 10 and 12 in a small range of positive values
of the electric field. For transitions labelled  with 7 and 10 this variation is a
consequence of an anticrossing effect around $F=100$ kV/cm, between the corresponding
energy levels $N_v = 5$ (in transition 7) and $N_v = 4$ (in transition 10) for the
$hh$-valence band with $m=0$ (see Table \ref{table2}). Similarly, transition 12 has a
sharp variation at $F\approx 150$ kV/cm because of an anticrossing of the
corresponding valence band energy level at this value of the electric field. Then, it
is possible to say that for some transitions the intensity of the dielectric signal
vanishes at certain values of the electric field and a given lens configuration,
leading to a non-monotonic behavior of the oscillator strengths as a function of the
electric field.

\begin{figure}
  \centerline{\epsfig{file=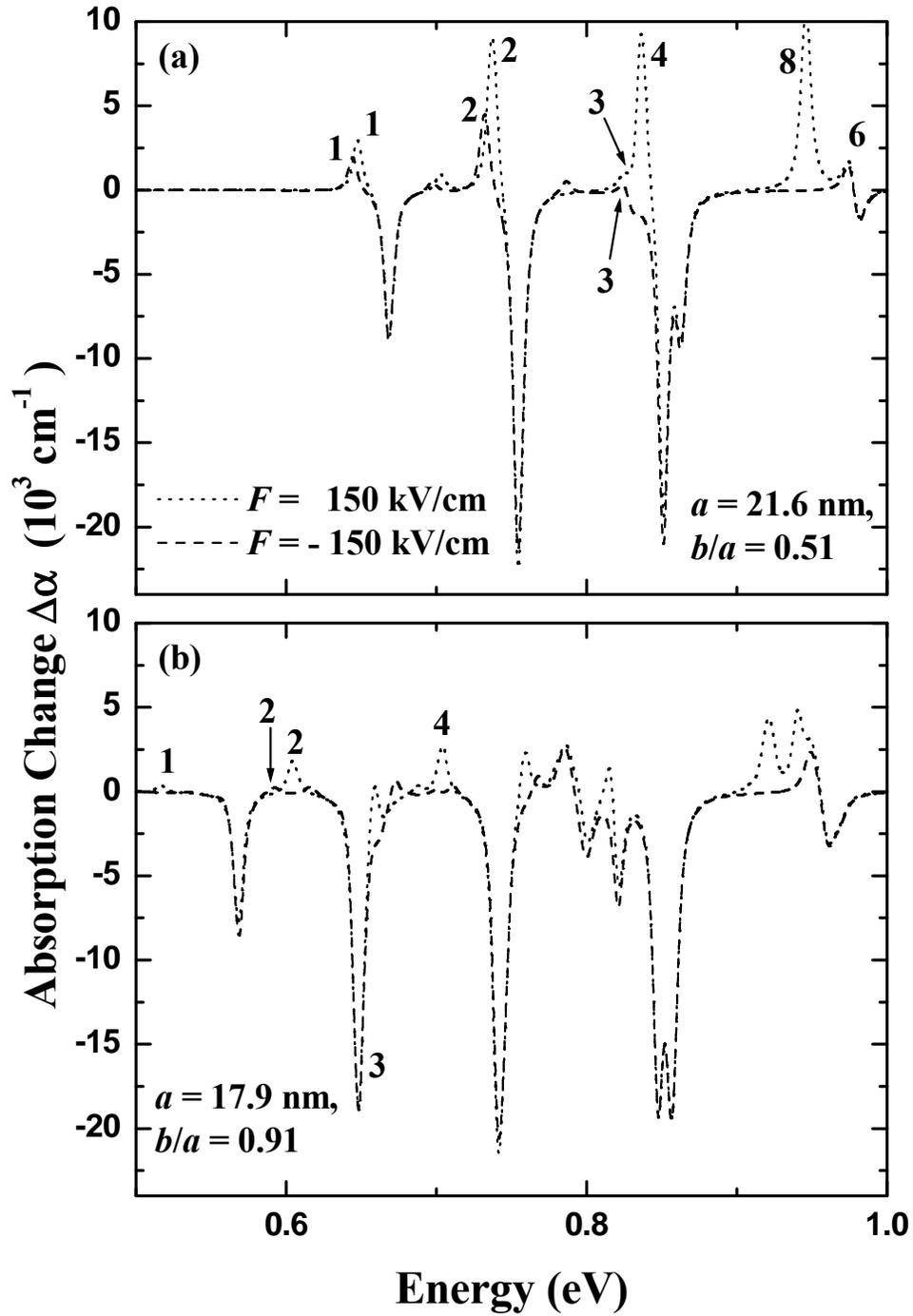,width=140mm, clip =}}
  \caption{Electroabsorption as a function of the energy. (a) Configuration corresponding to
  Fig. \ref{figInAsCD2} (a) and (b), (b) Configuration corresponding to
  Fig. \ref{figInAsCD2} (c).}
  \label{figInAsDAbs}
\end{figure}

\subsection{Electroabsorption.}

The optical properties of the semiconductor structure have led to the most powerful
techniques to study the electronic properties in solids \cite{dressel}. The
electroabsorption is one of these techniques used to inquire about the fundamental
properties of different quantum structures \cite{kenichiro}. The relation between the
imaginary part of the dielectric constant $\epsilon_2$ and the absorption coefficient
$\alpha$ is given by \cite{dressel}
\be
\epsilon_2(\omega) = \frac{n \, c}{\omega} \, \alpha(\omega)
\ee
where $n$ is the real part of the refractive index and $c$ the speed of light. We
obtained the real part of the refractive index according to
\be
n = \sqrt{\frac{\sqrt{\epsilon_1^2 + \epsilon_2^2} + \epsilon_1}{2}} + n_\infty
\ee
where $\epsilon_1$ and $\epsilon_2$ are the real and the imaginary part of the
dielectric constant, respectively, and $n_\infty$ is the high-frequency refractive
 index. For the case of an InAs SAQL we take $n_\infty = 3.517$ \cite{landolt}.
$\epsilon_1$ is calculated from $\epsilon_2$ and the Kramers-Kronig relations
\cite{PRB-DR}.

Previous studies \cite{kenichiro, henneberger} have used the absorption change
$\Delta \alpha(\omega,F) = \alpha(\omega,F) - \alpha(\omega,0) $ of a SAQL as a tool
to study low dimensional structures. In our case,  we study  the absorption change
for the two configurations used in Fig. \ref{figInAsCD2}. Different values of $\Delta
\alpha$ will be obtained according to $F$ being positive or negative. The results are
shown in Fig. \ref{figInAsDAbs}. In Fig. \ref{figInAsDAbs} (a) the lens configuration
of Fig. \ref{figInAsCD2} (a) and (b) has been used while in Fig. \ref{figInAsDAbs}
(b) the lens configuration corresponds to Fig. \ref{figInAsCD2} (c). In both cases,
the dotted (dashed) line represents $\Delta \alpha$  for $F=150 \; (F=-150)$ kV/cm.
Typical electroabsorption spectra are obtained and they depend on the direction
(positive or negative) of the electric field. A close correspondence is observed
among the peak positions in this figure and the transitions shown in the constant
dielectric (Fig. \ref{figInAsCD2}). Some transitions have been indicated for
comparison. $\Delta \alpha$ depends on the direction of the electric field due to the
spatial asymmetry of the lens, confirming that not only the magnitude but also the
direction of the field is important. Since this effect becomes important for stronger
confinement, the peak heights turn out to be larger as Fig. \ref{figInAsDAbs} (a)
shows. Thus, electroabsorption measurements would provide signatures of the
electronic structure of the SAQL.

\section{Conclusions.}

The present work has studied the influence of the lens asymmetry along the axial axis
on its electronic properties. The difference in the energy Stark-shift was
characterized with the parameter $\delta(F)$, which measures the separation of the
ground state energy as the electric field has a positive or negative direction along
the axial symmetry axis with same intensity. It was shown that the parameter
$\delta(F)$ increases for high values of the field and diminishes as the lens
deformation becomes larger (smaller ratio $b/a$). Nevertheless, even in this latter
case the eigenvalues of the lens domain can not be approximated by those of a
cylindrical domain with the same volume. Therefore, the full lens geometry must be
considered always to obtain the correct spectrum.

The consequences of this asymmetry on the dielectric constant of the SAQL were also
studied. We analyzed its behaviour as a function of  the lens deformation for a fixed
volume of the domain and a given absolute value of the field. Different transitions
are obtained depending on the lens deformation $b/a$, the radius $a$ and the
magnitude and direction of the external field. We emphasize the importance of the
direction of the electric field in the optical properties, such as the absorption
change, showing different behaviour according to a positive or negative direction
along the axis lens. Thus, in an experimental set up, an specific transition could be
tuned. Moreover, the correlation between the electroabsorption peaks (shown in this
work) and the electronic structure of the quantum lens (see details in Ref.
\cite{PRB-DR} and references therein) can be useful to characterize the geometrical
dimensions of these semiconductor nanostructures.

\section*{Acknowledgments}

This work was partially supported by Grant II 193-04/EXG/G (VIEP-BUAP). We also
thanks C. Trallero-Giner for clever discussions.

\bibliographystyle{prsty}
\bibliography{paper}

\begin{thebibliography}{10}

\bibitem{Yoffe2001}
A.~D. Yoffe, Advances in Physics {\bf 50},  1  (2001).

\bibitem{hawrylak-book}
L. Jacak, P. Hawrylak, and A. Wojs, {\em Quantum dots} (Springer-Verlag,
  Berlin, 1998).

\bibitem{Stranski-Krastanow}
I.~N. Stranski and L. Krastanow, Sitzungsber. Akad. Wiss. Wien {\bf 146},  797
  (1938).

\bibitem{fry00}
P.~W. Fry {\it et~al.}, Phys. Rev. Lett. {\bf 84},  733  (2000).

\bibitem{hawrylak96-1}
A. Wojs, P. Hawrylak, S. Fafard, and L. Jacak, Phys. Rev. B {\bf 54},  5604
  (1996).

\bibitem{jpacm}
A.~H. Rodr\'{\i}guez, C.~R. Handy, and C. Trallero-Giner, J. Phys.: Condens.
  Matter {\bf 15},  8465  (2003).

\bibitem{pss}
A.~H. Rodr\'{\i}guez and C. Trallero-Giner, Phys. Stat. Sol.(b) {\bf 230},  463
   (2002).

\bibitem{pss1}
A.~H. Rodr\'{\i}guez, L. Meza-Montes, C. Trallero-Giner, and S.~E. Ulloa, Phys.
  Stat. Sol. (b) {\bf 242},  1820  (2005).

\bibitem{C-T}
E. Casado and C. Trallero-Giner, Phys. Stat. Sol. (b) {\bf 196},  335  (1996).

\bibitem{PRB-DR}
A.~H. Rodr\'{\i}guez, C. Trallero-Giner, M. Mu{\~n}oz, and M.~C. Tamargo, Phys.
  Rev. B {\bf 72},  045304  (2005).

\bibitem{raymond98}
S. Raymond {\it et~al.}, Phys. Rev. B {\bf 58},  R13415  (0098).

\bibitem{jap}
A.~H. Rodr\'{\i}guez and C. Trallero-Giner, J. Appl. Phys. {\bf 95},  6192
  (2004).

\bibitem{emp99}
E. Men{\'e}ndez-Proupin, C. Trallero-Giner, and S.~E. Ulloa, Phys. Rev. B {\bf
  60},  16747  (1999).

\bibitem{martin}
M. Mu{\~n}oz {\it et~al.}, App. Phys. Lett. {\bf 83},  4399  (2003).

\bibitem{landolt}
L.-B. Tables, {\em Numerical Data and Functional Relationships in Science and
  Technology}, Vol.~37a of {\em Landolt-B{\"o}rnstein New Series, Group III},
  edited by {O}. {M}adelung ed. (Springer-Verlac, Berlin, 1982).

\bibitem{dressel}
M. Dressel and G. Gr{\"u}ner, {\em Electrodynamics of Solids: optical
  properties of electron in matter} (University Press, Cambridge, 2002).

\bibitem{kenichiro}
K. Tanaka, T. Takahashi, and T. Kondo, Phys. Rev. B {\bf 71},  045312  (2005).

\bibitem{henneberger}
F. Henneberger, S. Schmitt-Rink, and E.~O. G{\"o}del, {\em Optics of
  Semiconductor Nanostrutures} (Akademie Vlg., Berlin, 1993).

\end{thebibliography}

\end{document}